\documentclass[showpacs,preprintnumbers,amsmath,amssymb,preprint]{revtex4}
\usepackage{graphics}
\begin{document}

\title{Break-up fragments excitation and the freeze-out volume}

\small
\author{Ad. R. Raduta$^{1,2}$,
E. Bonnet$^{1}$, B. Borderie$^{1}$,
N. Le Neindre$^{1}$ and M. F. Rivet$^{1}$}
\address{$^{1}$Institut de Physique Nucleaire, IN2P3-CNRS, F-91406 Orsay
cedex, France\\
$^{2}$NIPNE, Bucharest-Magurele, POB-MG 6, Romania}

\begin{abstract}
We investigate, in microcanonical multifragmentation models,
the influence of the amount of energy dissipated in break-up fragments
excitation on freeze-out volume determination.
Assuming a limiting temperature decreasing with nuclear mass, we
obtain for the Xe+Sn at 32 MeV/nucleon reaction
[J. D. Frankland {\it et al.}, Nucl. Phys. A689, 905 (2001); A689, 940 (2001)]
a freeze-out volume almost half the one deduced using
a constant limiting temperature.
\end{abstract}
\maketitle

\section{Introduction}

One on the most provocative objectives of nowadays nuclear physics,
the determination of the phase diagram of excited nuclear systems,
requires precise measurements of relevant thermodynamical observables, 
volume and temperature.

From the theoretical point of view, the freeze-out volume is regarded
as an external constraint as in
microcanonical multifragmentation models \cite{mmmc,smm,randrup,mmm}
or simply as the volume in which are situated the non-interacting fragments
as in dynamical models \cite{parlog}.
From the experimental point of view,
the estimation of the freeze-out volume is delicate since break-up information
is altered by secondary particle emission and non-equilibrium phenomena.
The difficulty of determining the freeze-out volume is reflected in the
different values obtained by theoretical models
\cite{mmmc,smm,bob,indra3,mmm_indra}
and particle interferometry experiments \cite{aladin} which range from
2.5$V_0$ to 9$V_0$, where $V_0$ is the volume corresponding to
normal nuclear matter density.

The aim of the present study is to investigate to what extent the freeze-out volume
obtained by microcanonical multifragmentation models depends on the definition of
break-up fragments in terms of excitation energy.
Section II makes a review on how the observables characterizing the equilibrated state
of the system are determined by microcanonical multifragmentation models.
Section III presents the evolution of statistical and kinetic fragment distributions
of a given system when one modifies
the amount of energy dissipated as excitation energy of primary fragments.
Section IV presents the dramatic reduction of the freeze-out volume when a
mass-dependent limiting temperature is employed.
The conclusions are formulated in Section V.

\section{Fragment formation in microcanonical models}

Microcanonical models for multifragmentation describe the fragment partition
of the equilibrated source $(A, Z, E_{ex}, {\rm \bf L}, V)$ taking into account all
configurations that are geometrically possible and
are not forbidden by conservation laws.
Thus, the available energy of the system ($E_{ex}$) is used up in
fragment formation ($Q$), internal excitation ($\sum_i \epsilon_i$),
fragment-fragment Coulomb interaction  ($\sum_{i<j} V_{ij}$) and
thermal motion $K$,
\begin{equation}
K=E_{ex}-Q-\sum_i \epsilon_i-\sum_{i<j} V_{ij},
\label{eq:K}
\end{equation}
and determines together with fragments internal state the statistical weight
of any configuration \cite{mmm},
\begin{equation}
W_C \propto \frac1{N_C!}\prod_{n=1}^{N_C}\left(\Omega 
\frac{\rho_n(\epsilon_n)}{h^3}(mA_n)^{3/2}\right)\\
\cdot \frac{2\pi}{\Gamma(3/2(N_C-2))} \cdot \frac{1}{\sqrt{({\rm det} I})}
\cdot \frac{(2 \pi K)^{3/2N_C-4}}{(mA)^{3/2}}.
\label{eq:wc}
\end{equation}

From Eq. \ref{eq:K} it is obvious that there is a strong interplay between all
energetic degrees of freedom. While the Q-value depends exclusively
on fragment partition, the Coulomb energy depends on both fragment partition
and volume. The excitation energy absorbed by fragments is determined by
the maximum allowed excitation and level density.
As fragments definition at break-up depends on the underlying break-up scenario,
it is postulated by each model.
Thus, MMMC \cite{mmmc} considers primary fragments rather cold which decay
mainly by neutron evaporation.
On the other hand, SMM \cite {smm} allows fragments to excite to much higher energies
but their internal temperature is fixed by the translational one, as authors expect
to happen when thermodynamical equilibrium is achieved.
Finally, MMM \cite{mmm} assumes the fragment binding energy as the upper limit
for internal excitation.
Obviously, more excited the break-up fragments are, more abundant is the
secondary particle emission. 
Recent experimental data \cite{houdan}
based on relative velocity correlations between light charged
particles and fragments provide for the excitation energy of
primary fragments obtained in central collisions of Xe+Sn from 32 to 50 MeV/nucleon
values around 2-3 MeV/nucleon confirming the hypotheses of hot primary fragments.

The established method to identify the equilibrated source corresponding to
a given experimental multifragmentation reaction is to tune the 
values of the source parameters such as to describe as well as possible
all available statistical and kinetic fragment distributions.

Even if the conclusions of the present study hold for all versions of the
microcanonical multifragmentation model we shall use the MMM version
\cite{mmm,mmm_indra} to illustrate them.
In particular, we shall focus on the case of the
Xe+Sn at 32 MeV/nucleon multifragmentation reaction \cite{frankland}
because it offers one of the most complete set of experimental data in the literature
and made the object of a rather comprehensive investigation
on source identification using MMM \cite{mmm_indra}.

Thus, employing the level density formula,
\begin{equation}
\rho(\epsilon)=\frac{\sqrt{\pi}}{12 a^{1/4}\epsilon^{5/4}}
\cdot \exp(2 \sqrt{a \epsilon}) \cdot \exp(-\epsilon/\tau),
\label{eq:rho}
\end{equation}
with $a=0.114 A+0.098 A^{2/3}$ MeV$^{-1}$ \cite{iljinov}
and $\tau$=9 MeV,
Ref. \cite{mmm_indra} reached the conclusion that the freeze-out volume is 8 - 9 $V_0$.

To test the excitation energy effect of freeze-out volume identification,
we hold the fragments binding energy as the upper limit of their excitation
and modify the amount of energy dissipated in internal excitation by varying
the limiting temperature $\tau$.
Three values will be considered in our study: $\tau$=9 MeV
as in Ref. \cite{mmm_indra}, 
an arbitrarily chosen smaller value, 5 MeV and a mass dependent limiting temperature,
$\tau(A)=4.3+18.3/\sqrt{A}-19.9/A$ MeV.
The decrease of the limiting temperature with the fragment mass was demonstrated
by phenomenological analyses of the evolution of
the caloric curve plateau temperature and mean field calculations \cite{natowitz,baldo}.
The considered expression of $\tau(A)$ was obtained from a fit over data presented
in Refs. \cite{natowitz} and \cite{baldo} and ranges between the two above
considered values ($\tau(A=5)$=8.5 MeV and $\tau(A=200)$=5.5 MeV).

\section{Effect of fragment excitation on fragment distributions} 

\begin{figure}
\resizebox{0.6\textwidth}{!}{%
  \includegraphics{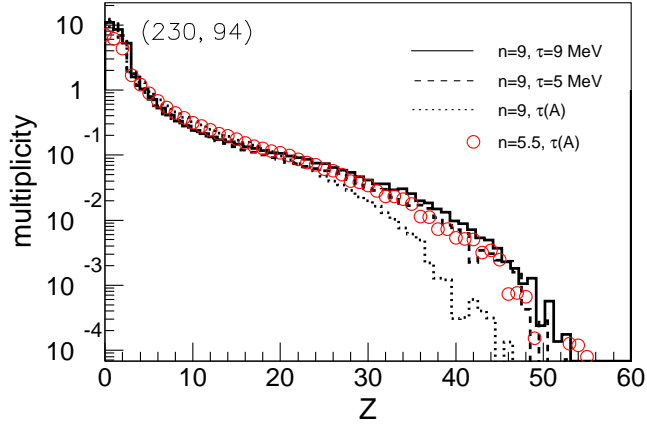}}  
\caption{Break-up charge distributions corresponding to the (230,94) nuclear system with 
$E_{ex}$=5.3 MeV/nucleon and 9$V_0$ (solid, dashed and dotted lines) and
5.5$V_0$ (open circles).
The amount of excitation energy absorbed by primary fragments is modified via the limiting
temperature, $\tau$.}
\label{fig:yz}
\end{figure}

Fig. \ref{fig:yz} depicts the break-up stage charge distributions obtained from
the multifragmentation of the (230, 94) nuclear system with $E_{ex}$=5.3 MeV/nucleon
and $V=9V_0$. This is the source for which, 
assuming the break-up fragments as hard non-overlapping spheres localized in
a spherical container (the freeze-out volume),
MMM obtained the best agreement with the
experimental data corresponding to Xe+Sn at 32 MeV/nucleon \cite{mmm_indra}.
As one may see, in this particular situation,
the effect of diminishing the limiting temperature from 9 to 5 MeV
is very small, $Y(Z)$ corresponding to lower excitation energies
being barely steeper thus indicating a more advanced fragmentation. Indeed, following the
evolution of average partial energies summarized in Table I,
it results that
the decrease of average total excitation energy leads to
an increase of almost 50\% of the Q-value.
However, the total number of fragments increases by only 18\% such that the average
total Coulomb field (which is maximum when the freeze-out volume is populated
uniformly) increases very little.
The energy excess is transformed into thermal motion.

\begin{table}
\begin{center}
\caption{Break-up stage values of average total Coulomb energy,
thermal kinetic energy, Q-value, excitation energy expressed in MeV and
total multiplicity (including neutrons and light charged particles)
obtained in the multifragmentation of (230, 94) nuclear system with
$E_{ex}$=5.3 MeV/nucleon and different freeze-out volumes and excitation of primary
fragments, as indicated in the first column.}
\begin{tabular}{|c|c|c|c|c|c|}
\hline
($V/V_0$, $\tau$)&$V_{Coulomb}$ & $K_{thermal}$ & $Q$ & $\epsilon_{tot}$ & Total multiplicity\\
\hline
(9, 9) & 390.61 & 287.67 & 186.59 & 354.13 & 32.5 \\
(9, 5) & 406.96 & 347.29 & 270.86 & 193.89 & 38.4 \\
(9, $\tau(A)$) & 413.33 & 294.01 & 232.56 & 279.10 & 34.0 \\
(5.5, $\tau(A)$) & 473.17 & 257.63 & 145.37 & 342.83 &26.2  \\
\hline
\end{tabular}
\end{center}
\label{table:en}
\end{table}

\begin{figure}
\resizebox{0.6\textwidth}{!}{%
  \includegraphics{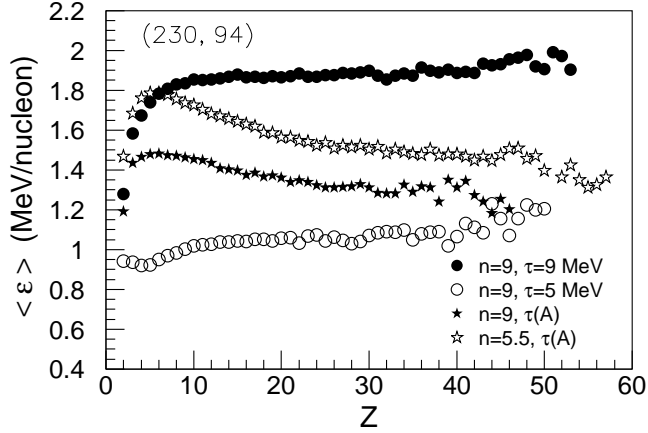}}  
\caption{Average excitation energies of primary fragments obtained from the
multifragmentation of the (230,94) nuclear system with 
$E_{ex}$=5.3 MeV/nucleon and different freeze-out volumes and limiting temperatures.}
\label{fig:ex_z}
\end{figure}

Much more exciting is the situation when the constant $\tau$ is replaced with
the mass-dependent one. As the produced fragments have $Z$ smaller that 60,
$\tau$ ranges between 6.3 and 8.5 MeV and the excitation energy is lower than
the one obtained for $\tau$=9 MeV and higher than the one corresponding to
$\tau$=5 MeV, as confirmed by the values in Table I 
and average excitation energy per nucleon of primary fragments represented
in Fig. \ref{fig:ex_z} as a function of fragment charge.
As expected, the evolution of average total excitation energy is reflected
in the evolution of average excitation of each emitted fragment.
A more interesting remark is that, as far as the limiting temperature is constant, 
the average excitation energy per nucleon increases slowly with fragment size
reflecting the increase of the binding energy over the considered mass domain,
while a mass decreasing $\tau$ produces an opposite behavior.
Moreover, the steep increase of the binding energies for small fragments corroborated with
the decreasing $\tau$ induces a hump in the fragment excitation energy distributions.
It is worthwhile to mention here that the distribution corresponding to
$\tau$=9 MeV is in acceptable agreement with the experimentally deduced one \cite{houdan}.

An intermediate value of excitation energy induces an intermediate Q-value and, not obvious,
a strong narrowing of the $Y(Z)$ distributions. The effect on fragment partitions can be
understood analyzing the statistical weight of a configuration.
The ratio between a partition with $N$ fragments $\{A_1, A_2, ... , A_i, ..., A_N\}$
and the partition in which the $i$-th fragment is replaced by two smaller fragments
$i'$ and $j'$ ($A_{i'}+A_{j'}=A_i$) is,
\begin{eqnarray}
\nonumber
\frac{W_{C_N}}{W_{C_{N+1}}} \propto \frac{\rho(\epsilon_{i})}
{\rho(\epsilon_{i'}) \rho(\epsilon_{j'})}
=
\left( \frac{a_{i'}a_{j'}}{a_i} \right)^{1/4} \cdot
\left( \frac{\epsilon_{i'} \epsilon_{j'}}{\epsilon_i} \right)^{5/4} \cdot
\frac{12}{\sqrt{\pi}} &\cdot&
\exp \left( 2 \sqrt{a_i \epsilon_i}- 2 \sqrt{a_{i'} \epsilon_{i'}}
- 2 \sqrt{a_{j'} \epsilon_{j'}}\right) \\
&\cdot& \exp\left( -\frac{\epsilon_i}{\tau_i}+\frac{\epsilon_{i'}}{\tau_{i'}}
+\frac{\epsilon_{j'}}{\tau_{j'}}\right).
\label{eq:rap_w}
\end{eqnarray}
To simplify the estimation one can consider that for the large fragments ($Z > 20$)
the excitation energy per mass number is constant, $\epsilon=<\epsilon> \cdot A$.
If $\tau$ is constant,
 the factor $\exp\left( -\frac{\epsilon_i}{\tau_i}+\frac{\epsilon_{i'}}{\tau_{i'}}
+\frac{\epsilon_{j'}}{\tau_{j'}}\right)$ vanishes
and the expression in Eq. \ref{eq:rap_w} has no explicit dependence on $\tau$.
In this case, the fragment partitions are selected
by the other factors entering the statistical weight formula, including those which depend
on the internal excitation ($\epsilon^{5/4} \cdot \exp(2\sqrt{A \epsilon})$).
If $\tau$ is decreasing with $A$, the last factor in Eq. \ref{eq:rap_w} is smaller than 1,
meaning that configurations with smaller fragments will be favored, as reflected by the
$Y(Z)$ distributions (see Fig. \ref{fig:yz}).

\begin{figure}
\resizebox{0.6\textwidth}{!}{%
  \includegraphics{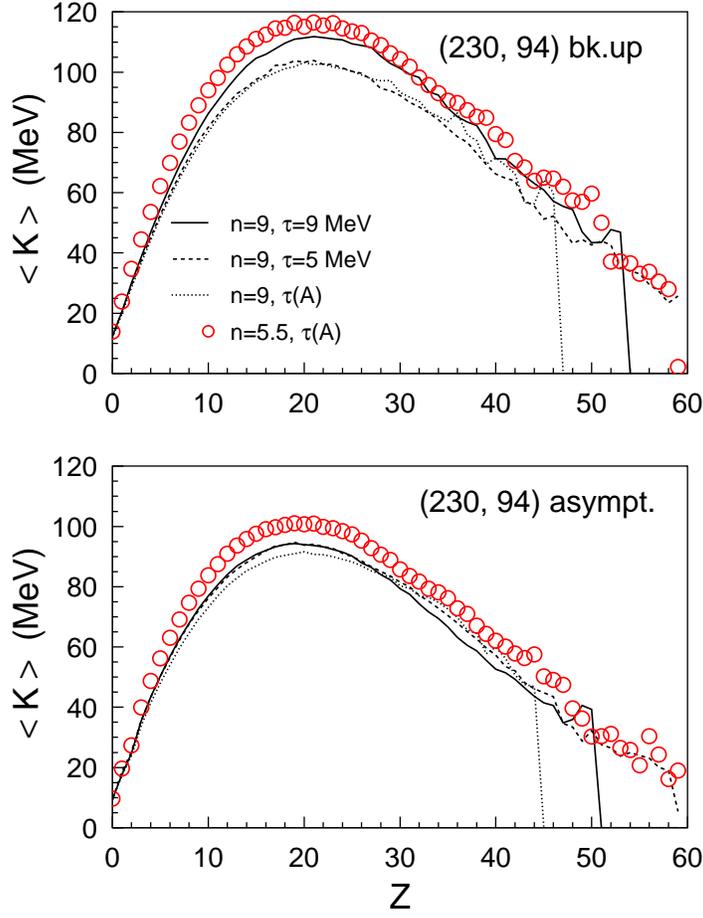}}  
\caption{Break-up stage (upper panel) and asymptotic stage (lower panel) mean kinetic energy
distributions as a function of charge obtained for the above considered cases.
}
\label{fig:kz}
\end{figure}

Because an additional key quantity
to determine the freeze-out volume is the fragment average kinetic
energy distribution as a function of charge,
in Fig. \ref{fig:kz} are presented the break-up and asymptotic stage results for the above discussed
situations. To maintain the comparison with the Xe+Sn at 32 MeV/nucleon case straightforward,
we considered for all cases the collective energy equal to 1.4 MeV/nucleon as
in Ref. \cite{mmm_indra}.
While the increase of both Coulomb and thermal energy obtained when $\tau$=9 MeV 
is replaced with 5 MeV would seem to produce higher $<K>$ versus $Z$
distributions, the upper panel of Fig. \ref{fig:kz} shows the opposite. This effect is due
to an even stronger increase of the total number of emitted fragments which automatically
results in smaller kinetic energy per fragment \cite{comment_viola}.
Concerning the evolution from $\tau$=5 MeV to $\tau(A)$, $<K>$ versus $Z$ is slightly affected
meaning that partial energy and total number of fragments modifications compensate each other.

Because the experimentally detected fragments are the asymptotic ones, the lower panel of  
Fig. \ref{fig:kz} depicts the asymptotic $<K>$ versus $Z$ distributions. More excited the fragments
are, stronger will be the effect of particle evaporation.
Thus, despite the average kinetic energies of intermediate size fragments obtained for 
$\tau$=9 MeV is 10 MeV larger than the ones corresponding to $\tau$=5 MeV, after sequential
evaporations they overlap.
 
To recover the initial fragment yield distribution with a $\tau(A)$ dependence,
one has to diminish the freeze-out volume.
In Figs. \ref{fig:yz} and \ref{fig:kz} are represented with open circles
the corresponding distributions obtained for the same value of $E_{ex}$=5.3 MeV/nucleon.
As indicated in the figure, $Y(Z)$ gets its initial shape for $V=5.5V_0$.
Such a small freeze-out volume results in a stronger Coulomb field 
(Table I)
moving upward the $<K>(Z)$ distribution.
This suggests that in this case the description of the experimental data will require a smaller
amount of flow.

\section{Application on experimental data}

To illustrate quantitatively the impact of assuming lower fragment excitation energies
on freeze-out volume determination we choose the most spectacular situation
of a mass dependent $\tau$.
To characterize the complete state of the fused system we adopt the
procedure used in Ref. \cite{mmm_indra} and the same Xe+Sn at 32 MeV/nucleon reaction. 
The freeze-out volume and source excitation are mainly determined 
by fitting the $Y(Z)$ distribution, while the source size and excitation are fixed by
the bound charge multiplicity and multiplicity of intermediate mass fragments. 
The amount of flow and the flow profile $\alpha$
are tuned such as to obtain a good agreement with the experimental average kinetic energy
distribution as a function of charge. Thus, the newly identified source is:
(220, 88), $E_{ex}$=5.1 MeV/nucleon, $V=5V_0$, $E_{flow}$=1 MeV/nucleon, $\alpha=1.8$.

\begin{figure}
\resizebox{0.6\textwidth}{!}{%
  \includegraphics{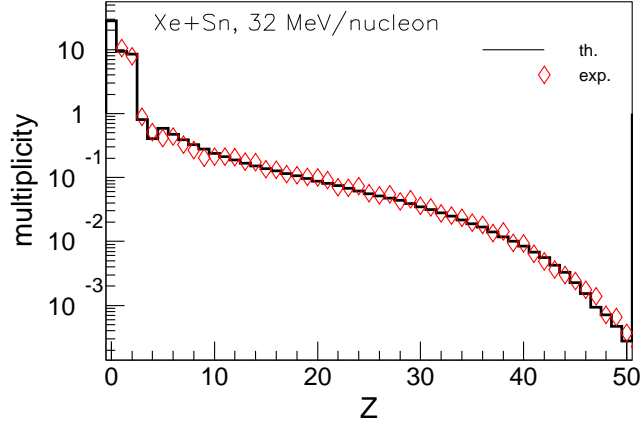}}  
\caption{Calculated asymptotic stage (histogram) and experimental (open symbols)
charge distributions corresponding to Xe+Sn at 32 MeV/nucleon.}
\label{fig:yz_220+exp}
\end{figure}

\begin{figure}
\resizebox{0.6\textwidth}{!}{%
  \includegraphics{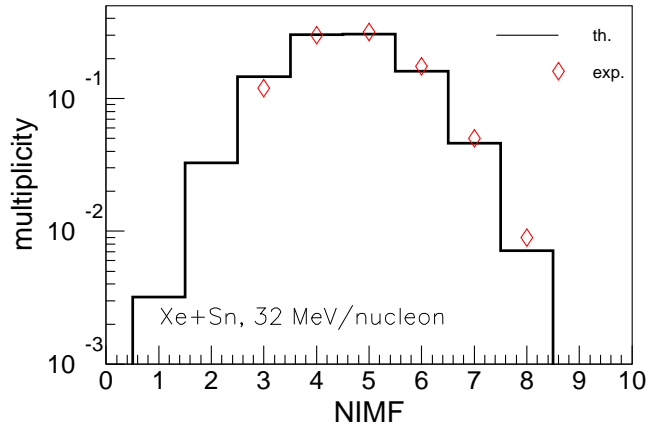}}  
\caption{Calculated asymptotic stage (histogram) and experimental (open symbols) 
distributions of multiplicity of number of intermediate mass fragments ($Z \ge 5$)
corresponding to Xe+Sn at 32 MeV/nucleon.}
\label{fig:imf_220+exp}
\end{figure}

\begin{figure}
\resizebox{0.6\textwidth}{!}{%
  \includegraphics{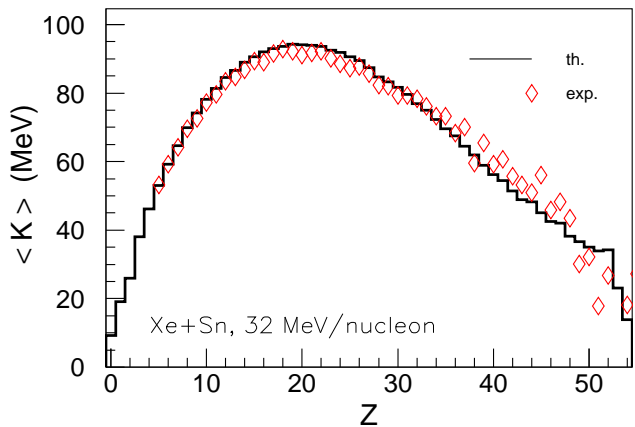}}  
\caption{Calculated asymptotic stage (histogram) and experimental (open symbols) 
distributions of fragment average kinetic energy
as a function of charge corresponding to Xe+Sn at 32 MeV/nucleon.}
\label{fig:kz_220+exp}
\end{figure}

Figs. \ref{fig:yz_220+exp}, \ref{fig:imf_220+exp} and \ref{fig:kz_220+exp} 
present the calculated asymptotic stage
charge multiplicity distribution, multiplicity of number of intermediate mass fragments
($Z \geq 5$)
and, respectively, average fragment kinetic energy as a function of charge
in comparison with the experimental data.

The most striking result is the reduction of the freeze-out volume to about half of the previously
determined value, $9V_0$. The modifications on mass and charge are minor (6\%).
The decrease of 30\% of the added flow is a consequence of the stronger Coulomb field
produced by the compressed matter for the same fragment partition.

The proposed set of observables is not expected to be necessarily the real one since
the values depend dramatically on the assumed limiting temperature, but are perfect examples to
illustrate how break-up hypotheses dependent are the results obtained by
statistical multifragmentation models. 

\section{Conclusions}

We investigate, in nuclear multifragmentation models,
the sensitivity of freeze-out volume determination
on primary fragment definition in terms of excitation energy.
Assuming a mass dependent limiting temperature in concordance with recent phenomenological estimations
and theoretical calculations, we distinguish a strong narrowing of charge distributions so that,
to recover the agreement with a witness distribution, we have to assume a freeze-out volume
almost half the one obtained with a constant limiting temperature.

Given the importance of accurate determination of the freeze-out volume,
model-independent methods to measure this quantity
and additional experimental information of the excitation state of primary fragments
are necessary.

\end{document}